\documentclass[12pt,preprint]{aastex}

\usepackage{subfigure}
\usepackage{cases}
\usepackage{graphicx}
\usepackage{natbib}
\bibpunct{(}{)}{;}{a}{}{,} 
\usepackage{txfonts}
\usepackage{graphics}
\usepackage{graphicx}
\newcommand{\be}{\begin{equation}}
\newcommand{\ee}{\end{equation}}

\slugcomment{Not to appear in Nonlearned J., 45.}

\shortauthors{fan et al.}

\begin{document}


\title{The dust properties of $z\sim3$ MIPS-LBGs  from photo-chemical models}


\author{X.L. Fan\altaffilmark{1,2,3,4}, A. Pipino\altaffilmark{5} and F. Matteucci\altaffilmark{3,4,6} }



\altaffiltext{1}{fan@oats.inaf.it}
\altaffiltext{2}{Hubei University of Education, 430205, Wuhan, Hubei, China}
\altaffiltext{3}{Dipartimento di Fisica, Sezione di Astronomia, Universit\`a di Trieste, via G.B. Tiepolo 11, I-34131, Trieste, Italy}
\altaffiltext{4}{I.N.A.F. Osservatorio Astronomico di
 Trieste, via G.B. Tiepolo 11, I-34131, Trieste, Italy}
\altaffiltext{5}{Institut fur Astronomie, ETH Zurich, 8093 Zurich, CH}
\altaffiltext{6}{I.N.F.N., Trieste, Via Valerio 2, 34100, Trieste, Italy}


\begin{abstract}
The stacked spectral energy distribution (SED) of  Multiband Imaging Photometer for Spitzer (MIPS) 24$\mu m$ detected Lyman break galaxies (MIPS-LBGs) is fitted by means of the spectro-photometric  model GRASIL with an ``educated'' fitting approach which benefits from the results of chemical evolution models. The star formation rate-age-metallicity degeneracies of SED modelling are broken  by using the star formation history and chemical enrichment history suggested by chemical models. The dust mass, dust abundance and chemical pattern of  elements locked in the dust component are also directly provided by chemical models. Using our  new ``fitting" approach,  we derive the total mass $M_{tot}$, stellar mass $M_{\ast}$, gas mass $M_{g}$, dust mass $M_{d}$, age and star formation rate (SFR) of the  stacked MIPS-LBG in a self-consistent way. Our estimate of $M_{\ast}= 8\times 10^{10}$ of the stacked MIPS-LBG agrees with other works based on UV-optical SED fitting.  We suggest that the MIPS-LBGs at $z\sim3$ are young (0.3-0.6 Gyr), massive ($M_{tot} \sim 10^{11} M_{\odot}$), dusty ($M_{d} \sim 10^{8} M_{\odot}$), metal rich ( $Z \sim Z_{\odot} $) progenitors of elliptical galaxies suffering a strong burst of star formation (SFR $\sim 200 M_{\odot}/yr$ ). Our estimate of $M_{d}=7 \times 10^{7} M_{\odot}$  of the  stacked MIPS-LBG  is about a factor of eight lower than the estimated value  based on single temperature grey-body fitting,  suggesting that  self-consistent SED models are needed to estimate  dust mass. By comparing with the Milky Way molecular cloud and dust properties, we suggest that denser and dustier environments and flatter dust size distribution are likely in high redshift massive star forming galaxies.  These dust properties, as well as the different types of SFHs,  can cause different SED shapes between high redshift star-forming ellipticals and local star-burst templates. This discrepancy of SED shapes could in turn explain the non detection at submillimeter wavelengths, of IR luminous ($L_{IR} \succeq 10^{12} L_{\odot} $) MIPS-LBGs.
\end{abstract}





\section{Introduction}
The last decade witnessed a tremendous increase in the knowledge
of high redshift star forming galaxies. Large samples,
obtained with different selection techniques \cite[see][for a review]{Shapley2011ARA&A..49..525S},
yielded statistical information about masses \cite[e.g.][]{Erb2006ApJ...646..107E,Daddi2007ApJ...670..156D}, star formation rates and dust attenuation \cite[e.g.][]{Hopkins2006ApJ...651..142H,Reddy2008ApJS..175...48R,Reddy2010ApJ...712.1070R},
as well as metallicities \cite[e.g.][]{Erb2006ApJ...644..813E} for $z\simeq 2$ galaxies.

In a small subset of well studied objects, a more accurate determination
of their chemical abundance pattern, dust depletion, kinematics, was possible,
due to lensing magnification \citep[e.g.][]{Hainline2009ApJ...701...52H} and deep integral field observations \citep[e.g.][]{Law2012arXiv1206.6889L,Gnerucci2011A&A...528A..88G,Genzel2011ApJ...733..101G}.

Taken together, these observables shaped our view of $z=2-3$ redshift star forming galaxies
as relatively evolved systems,  as witnessed by their metallicities \citep[e.g.][]{Erb2006ApJ...644..813E,Calura2009A&A...504..373C}, whose star formation rates are high due to the availability of large gas reservoirs
as opposed to mergers \citep[e.g.][]{Genzel2010MNRAS.407.2091G,Rodighiero2011ApJ...739L..40R,Kaviraj2012MNRAS.tmpL..26K}.

Among the various methods, the Lyman Break Technique \citep{Steidel1993AJ....105.2017S} was particularly successful in yielding
insight into high redshift star forming galaxies.
Lyman break galaxies (LBGs) are observed at different redshifts: z$\sim$1 \citep{Burgarella2007MNRAS.380..986B}, z$\sim$ 1.4-2.5 ``BX" and
``BM" \citep{Steidel2004ApJ...604..534S}, z $\sim$3 \citep{Steidel2003ApJ...592..728S}, z $\sim$ 4,5,6 \citep{Vanzella2009ApJ...695.1163V}
by means of the
photometrical selection technique for rest-frame 912 $\AA$ Lyman-continuum discontinuity in a desired redshift interval. %
To infer the properties of stars, interstellar medium (ISM) and dust,  LBGs have been studied
in terms of stellar light,
interstellar absorption lines and rest-frame UV continuum  at $ z\gtrsim  3  $ \citep{Steidel2003ApJ...592..728S,Vijh2003ApJ...587..533V}.
One of the best studied LBGs is the gravitational lensed LBG MS 1512-cB58  \,(hereafter cB58) at redshift $z = 2.7276$, which seems to be  a young  system
undergoing rapid star  formation and possibly destined to be an elliptical galaxy \citep{Pettini2002ApJ...569..742P}.
By means of cosmological hydrodynamical simulations, \cite{Dave2000ASPC..200..173D}  suggested that  LBGs at z $\sim$ 3 are the most massive galaxies at that epoch without ruling out the merger-induced star-burst galaxies contribution.  Using a  chemodynamical model, \cite{Friaca1999MNRAS.305...90F} suggested that LBGs at z $\sim$ 3 are the progenitors of present-day massive spheroids. \cite{Matteucci2002lbg} and \cite{Pipino2011A&A...525A..61P} suggested that LBGs at z $\sim$ 3 can be young small  star forming elliptical galaxies by fitting the abundance pattern of cB58 by means of galactic chemical evolution models. The stellar population properties of LBGs  at z $\sim$ 3 are studied by \cite{Sawicki1998AJ....115.1329S,Papovich2001ApJ...559..620P,Magdis2008MNRAS.386...11M,Magdis2010MNRAS.401.1521M,Pentericci2010A&A...514A..64P}.

In the absence of accurate spectroscopic data
available for large galaxy  samples, the stellar properties, such as the age and stellar mass are estimated by spectral energy distribution (SED) fitting based on stellar population synthesis models  where star formation history (SFH), metallicity and dust extinction are independent free parameters.   Whilst some studies adopt a full treatment of  radiative transfer  \citep[e.g.][]{Takagi2003MNRAS.340..813T,Siebenmorgen2007A&A...461..445S,Rowan-Robinson2012IAUS..284..446R,LoFaro2013ApJ...762..108L},  in most cases empirical attenuation laws are  usually  adopted to deconvolve the effect of the dust from the SED. This means that molecular cloud (MC) condition, and dust properties, such as dust mass, dust abundance and dust size distribution, are not directly studied.   Only a few works attempt to fit the SEDs of LBGs from the UV to the IR \citep[e.g.][]{Magdis2010ApJ...720L.185M,Magdis2011A&A...534A..15M}.  However, separate models are adopted in these works to fit the UV-optical and the IR parts of the SED.  Furthermore, the dust temperature and mass  are usually estimated by a grey body fitting, and the IR luminosity are usually estimated with template SEDs from an observational  library. Whilst these assumptions might be justified as a necessary to
explore \emph{uncharted territory} and certainly provide
a good zero order estimate of the relevant physical quantities (masses, SFRs), we believe
that more sophisticated models should be adopted to understand the physical processes
acting in high redshift star forming galaxies at a finer degree of detail.

To this purpose, in previous work (Matteucci \& Pipino, 2001, Pipino et al., 2011),  we focussed on the chemical abundance pattern of a well studied single LBG cB 58 \citep{Pettini2000ApJ...528...96P}  and LBGs  from two surveys: AMAZE \citep{Maiolino2008A&A...488..463M} and LSD \citep{Mannucci2009MNRAS.398.1915M}. We self-consistently derived the SFH and the dust properties for these high redshift LBGs.
Here we want to address in detail the dust properties of $z\sim 3$ LBGs. In particular, we  will make use of the stacked SED of  Multiband Imaging Photometer (MIPS) 24$\mu m$ detected LBGs (MIPS-LBGs) at $z\sim3$  \citep{Magdis2010ApJ...720L.185M},  since they allow for the first time an entire coverage from the UV to the radio wavelenghts. MIPS-LBGs, per se, are not a special class of objects. However they are likely the most massive, most dust rich, most rigorously star forming LBGs  at redshift $z \sim 3$,   following a SFR-stellar mass relation  with a similar slope to that found at redshift z=1 \citep{Elbaz2007A&A...468...33E} and z=2 \citep{Daddi2008ApJ...673L..21D}, but a higher normalization factor \citep{Magdis2010MNRAS.401.1521M}. The inferred properties of the stacked MIPS-LBG SED  will represent the average massive LBGs  at redshift z $\sim$ 3. The high SFR and infrared luminosity suggest that MIPS-LBGs  should be detected at submm wavelengths.  However, this occurs in a small fraction of cases \citep{Chapman2000MNRAS.319..318C,Shim2007ApJ...669..749S,Chapman2009MNRAS.398.1615C}. This non detection far-IR counterparts of MIPS-LBGs could be caused by higher dust temperature, different dust spatial distribution and  lower SFR in LBGs  than in submillimetere galaxies (SMGs) \citep[see discussion in ][]{Chapman2000MNRAS.319..318C,Rigopoulou2006ApJ...648...81R,Magdis2010ApJ...720L.185M}.  The self-consistent SED model combined with chemical model will allow us to investigate this issue in more detail.

In our approach (see details below), we study the properties of  stellar population and ISM  by the spectro-photometric model GRASIL  \citep{Silva1998ApJ...509..103S}. Parameter degeneracy  is  a problem for any galactic SED fitting \citep[see reviews by][]{Gawiser2009NewAR..53...50G,Walcher2011Ap&SS.331....1W}. Therefore, our self-consistent chemical evolution  model \citep{Pipino2011A&A...525A..61P},  which can reproduce most of the chemical properties of ellipticals of different mass,  is adopted to  reduce the degree of freedom as much as possible in our SED modelling by GRASIL.  The combination of chemical evolution  models and   GRASIL  has been already adopted by \cite{Schurer2009MNRAS.394.2001S}  to model SEDs of different morphological type galaxies using the chemical and dust evolution models from  \cite{Calura2008A&A...479..669C}.

In this article, we aim at: 1) deriving the dust mass, and 2) investigating dust composition, temperature and grain size distribution of MIPS-LBGs.  We will also assess the robustness of previously derived galaxy masses and SFR, compare MIPS-LBGs with other  $z \sim 3$ LBGs,  confirm the nature of LBGs at z $\sim$ 3 from previous chemical models,   and try to solve the challenge of non detection far-IR counterparts of MIPS-LBGs.  The paper is organized as follows: in Sect. \ref{models}, we describe our models and fitting approach; our results and discussions are presented in Sect. \ref{result}; and our conclusions are drawn in Sect. \ref{summary}. Throughout the paper, we adopt a (0.7, 0.3, 0.7) cosmology.

\section{The models}\label{models}

We combine  the chemical evolution models for elliptical galaxies \citep{Pipino2011A&A...525A..61P} and the spectro-photometric  model GRASIL \citep{Silva1998ApJ...509..103S} to model the SED of the  stacked MIPS-LBG. The combination approach is fully described  in \cite{Schurer2009MNRAS.394.2001S} for galaxies of different morphological type.
 We direct the reader to  above  articles for equations and details. For the sake of convenience the two models will be briefly summarized in the next sections. In particular, the details of the our  ``fitting" approach will be described.
\subsection{The chemical evolution model}\label{chem_model}
The chemical abundance ratios versus metallicity are amongst the best indicators to constrain the SFH of a galaxy. However, there is no chemical data of the average MIPS-LBGs at z $\sim$ 3 to constrain the SFH. We adopt the SFHs of elliptical  models \citep{Pipino2011A&A...525A..61P}, which have  successfully reproduced the chemical abundance properties of normal ellipticals (see Pipino \& Matteucci, 2004) as well as those of low mass LBGs at z $\sim$ 3
\citep[see][for detail]{Pipino2011A&A...525A..61P}. In  this way, the SFH, metallicity enrichment and dust evolution history of a galaxy are  obtained from a self-consistent chemical evolution model.
The impact of this assumption is reviewed in the discussion section.

In these models, both the infall and the star formation timescale are assumed to decrease with galactic mass, in order to reproduce the ``chemical downsizing'' \citep[e.g.][]{Thomas2005ApJ...621..673T}.
The initial conditions  for  ellipticals allow for the formation by either collapse of a gas cloud  into the potential well of a dark matter halo or, more realistically, by the merging of several gas clouds. In any case, the timescale for both processes should be shorter than 0.5 Gyr, so that the ellipticals form very rapidly. The rapid gas assembly triggers an intense and rapid star-formation process that lasts until a galactic wind, powered by the thermal energy injected by stellar winds and  SNe (Ia, II) explosions, occurs. At that time, the thermal energy is equal or larger than the binding energy of
gas, and all the residual gas is assumed to be lost.  After the wind, the star formation stops and the galaxies evolve passively.
The evolution of the global metallicity as well as of 21 single chemical elements are studied in detail. 
Those elements are produced by single low and intermediate mass stars (AGB stars), SNe Ia, single massive stars (SNe II) and Type Ia SNe( white dwarfs in binary systems). 
The yields used in this paper are as follows: 1) For  single low and intermediate mass stars ($0.8 \leq M/M_{\odot}\leq 8$) we make use of the yields by \cite{van1997A&AS..123..305V}  as a function of metallicity; 2) For SNe Ia and SNe II we adopt the empirical yields by \cite{Francois2004A&A...421..613F}. These yields are a revised version of the \cite{Woosley1995ApJS..101..181W} (for SNe II) and \cite{Iwamoto1999ApJS..125..439I} (for SNe Ia) calculations adjusted to best fit the chemical abundances in the Milky Way. We consider that the dust producers are AGB stars, SNe Ia and SNe II. We also take in to account the dust  accretion and dust destruction processes in the ISM. Only the main refractory elements, C, O, Mg, Si, S, Ca, Fe, are depleted into dust, and we assume that stars can produce two different types of grains: i) silicate dust, composed of O, Mg, Si, S, Ca and Fe; and ii) carbon dust (graphite in the GRASIL language), composed of C. Since  there is no AGN signature in MIPS-LBGs  rest-frame UV
spectra as well as in their SED, we do not include QSO dust in this paper.

\begin{table*}
\begin{minipage}{120mm}
\scriptsize
\begin{flushleft}
\caption{Chemical model parameters and results.}\label{chem_p}
\begin{tabular}{l|lllll|ll|l}
\hline
\hline
model          &                &           &              &                  &             &             &                                        \\
name           & $M_{lum}$  	&$R_{eff}$  &  $\nu$ 	   & $\tau$            & $t_{gw}$    & SW &   SNe Ia and SN II& dust elements  \\
               &({$M_{\odot}$}) & ({kpc}) &  ({$Gyr^{-1}$})& (Gyr)& (Gyr)                     &   yields  & yields &                  \\
\hline
M310             &$3\times10^{10}$       &2        & 3            & 0.5   &  0.8    & 	 V$\&$G   & Fran\c cois&    C, Si, Fe, Mg, O ,(S, Ca)   \\
M11             &$10^{11}$       & 3       & 10            & 0.4    &  0.7    & 	  V$\&$G  & Fran\c cois&    C, Si, Fe, Mg, O, (S, Ca)   \\
M511             &$5\times10^{11}$       & 6       & 15            & 0.3   &  0.7    & 	V$\&$G    & Fran\c cois&    C, Si, Fe, Mg, O,  (S, Ca)   \\
\hline
\end{tabular}
\end{flushleft}
$M_{lum}$: the total mass of the galaxy at galactic wind time $t_{gw}$.\\
$R_{eff}$: the effective radius of the galaxy.\\
$\nu$: the star formation efficiency.\\
$\tau$: the infall time scale.\\
$t_{gw}$: the galactic wind time.  It is determined only by $M_{lum}$ (and hence $\nu$ and $\tau$). Note that it is shorter than in \cite{Pipino2011A&A...525A..61P}. This is because the $t_{gw}$ refers to the central region of the galaxy in \cite{Pipino2011A&A...525A..61P}, while it refers to the whole galaxy in this work. This is a  feature of  the outside-in formation formation scenarios \citep{Pipino2004MNRAS.347..968P}. \\
Single low and intermediate mass stars yields (SW yields) are from \cite{van1997A&AS..123..305V} ($V\&G$).\\
SNe Ia and  SNe II yields are from \cite{Francois2004A&A...421..613F}.\\
C, Si, Fe, Mg, O are the dust elements adopted in GRASIL, while  S, Ca are also the dust elements in chemical models.
\end{minipage}
\end{table*}
\begin{figure}
\begin{center}
\includegraphics[width=3in]{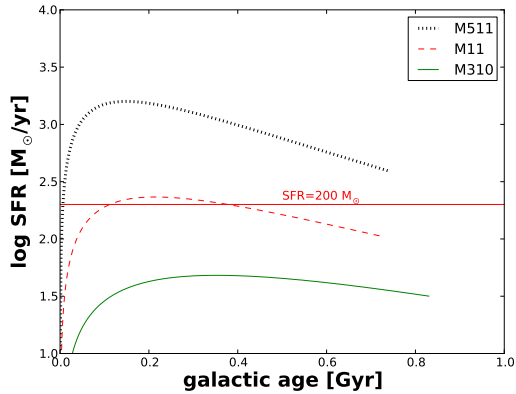}\\
\caption{The SFR versus time in different mass galaxies. M511, M11 and M310 represent $5\times 10^{11}, 10^{11}$ and $3\times 10^{10}M_{\odot}$ galaxies, respectively. The red horizontal line corresponds to a SFR=$200 M_{\odot}/yr$, which is the estimated value for the  average MIPS-LBGs.}\label{sed_sfr}
\end{center}
\end{figure}

\begin{figure}\label{che-inform}
\begin{center}
\subfigure[Total (red solid line), graphite (black dot line) and silicate (blue dashed line)  dust mass of $10^{11}M_{\odot}$ galaxy as a function of galactic age. ]{
 \label{che-inform1}
\includegraphics[width=3in]{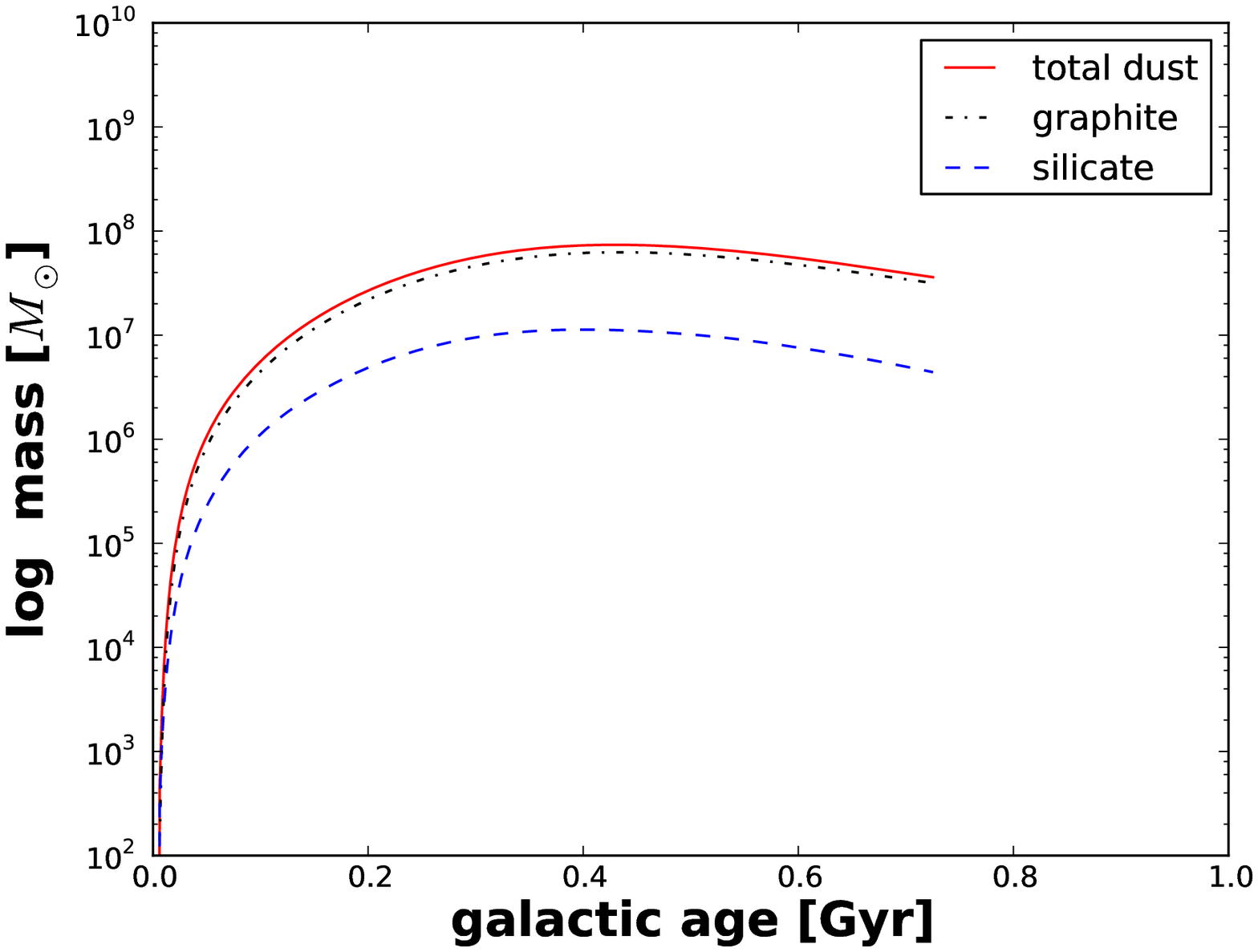}}
\subfigure[Dust to gas ratio of $10^{11}M_{\odot}$ galaxy as a function of galactic age.]{
 \label{che-inform2}
\includegraphics[width=3in]{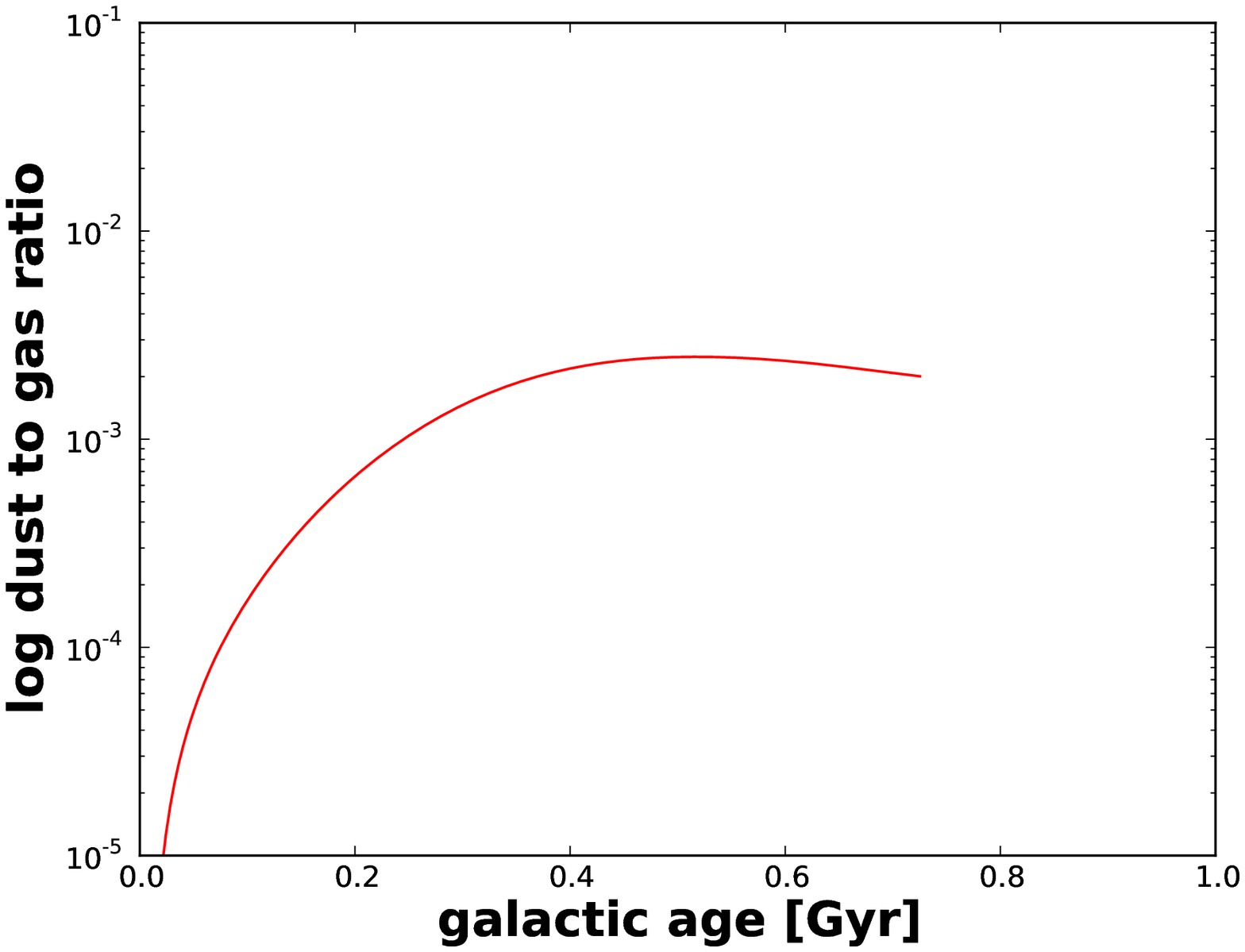}}
\subfigure[Metallicity of $10^{11}M_{\odot}$ galaxy as a function of galactic age.]{
 \label{che-inform3}
\includegraphics[width=3in]{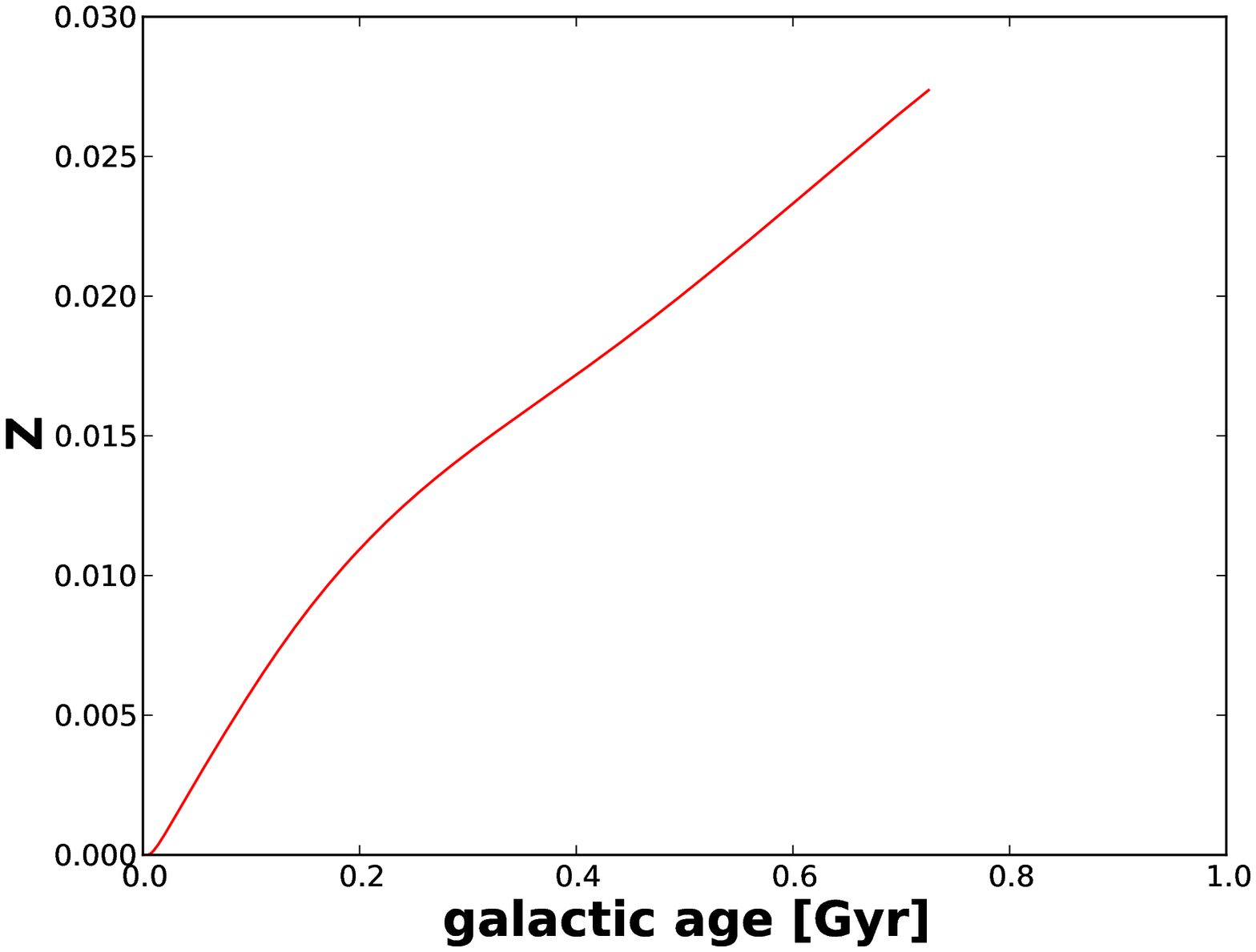}}
\caption{Output of the chemical evolution models adopted by GRASIL.  Values in the star forming phase  are shown, since we only focus on this phase. }
\end{center}
\end{figure}

\subsection{The spectro-photometric model}
Using the SFH,  chemical and dust evolution of a galaxy extracted from the chemical evolution model (e.g. as shown in Fig\ref{sed_sfr} and  Fig. \ref{che-inform1}-\ref{che-inform3}), we synthesize the SED of that galaxy with the spectro-photometric code GRASIL \citep{Silva1998ApJ...509..103S}. Briefly, emissions form stellar populations are calculated by evolutionary population synthesis  technique \citep{Bressan1994ApJS...94...63B}  using the chemical evolution model and the Padova simple stellar population model \citep{Bertelli1994A&AS..106..275B}. Younger stellar generations are more affected by dust obscuration in their birth place (molecular clouds) than older ones in diffuse ISM.   The effect of the age selective extinction of stellar populations, for the first time,  was modeled by a parametric approach adopting the parameter $t_{esc}$  in GRASIL \citep{Silva1998ApJ...509..103S}. This is
described in GRASIL assuming that the fraction of starlight radiated inside the clouds by stars is a function of the star age.
In practice, if $t_{esc}$ is the timescale for the process, $100\%$ of the stars younger than $t_{esc}$ are considered to radiate inside the MCs, and this percentage goes linearly to $0\%$ in 2$t_{esc}$.  The dust composition  in GRASIL  consists of graphite and silicate grains, and Polycyclic Aromatic Hydrocarbons (PAH) molecules, with a distribution of grain sizes for each composition.  The optical properties of
silicate and graphite grains are taken from \cite{Laor1993ApJ...402..441L}. The optical properties of PAH molecules are taken  from \cite{Draine2007ApJ...657..810D}.
Dust abundance and  dust composition relative abundances are obtained directly from the chemical evolution model.
For the molecular clouds a full radiative transfer calculation  is performed. The radiative transfer of starlight through dust is computed
along the required line of sight yielding the emerging SED \citep{Granato1994MNRAS.268..235G} based on lambda-iteration numerical method \citep{Efstathiou1990MNRAS.245..275E,Collison1991ApJ...368..545C}.  However for the diffuse dust, the effects of scattering are only approximated  by assuming an effective optical depth related to the true absorption and scattering optical depths by: $\tau_{eff} = [\tau_{abs} (\tau_{abs} + \tau_{sca})]^{1/2}$ and assuming that there is no re-absorption of the radiation emitted by the cirrus.

\subsection{The difference between  the chemical and spectro-photometric model}
The adopted chemical evolution models contain only one gas phase whereas GRASIL takes into account two gas phases: the molecular clouds and diffuse gas. We therefore assume that the chemical and dust-to-gas ratio and dust abundances are the same for those two gas phases. Ca and S are not contained  in GRASIL's dust composition. However, those two elements give a negligible contribution to dust-to-gas ratio. In this paper, we will only  consider the following refractory elements: C, Si, Fe, Mg and O. That is, quantities as the total dust mass will be computed
neglecting other elements (see \ref{che-inform2}). The effect caused by  the difference between ISM (ISM$\equiv$ gas + dust) metallicity and gas phase metallicity  on SED is also negligible. The  Salpeter IMF  \citep{Salpeter1955ApJ...121..161S} is adopted  in the two models.
The baryonic matter (i.e. stars plus gas) is assumed to follow the \cite{Jaffe1983MNRAS.202..995J} spatial  distribution in the chemical evolution model, while GRASIL adopts the King profile \citep{King1966AJ.....71...64K} spatial  distribution. The spatial distribution,  which determines the potential well,  only directly affects the galactic wind time in  chemical evolution models  (see discussion in section \ref{chem_model}).  The  geometry effect is not important  in the  elliptical galaxy  SED  \citep[e.g. see][for detail]{Silva1998ApJ...509..103S,Piovan2006MNRAS.370.1454P}.
 The effects of various parameters on the SED will be discussed in further work in detail. In this paper we only describe the general effects of parameters on deriving physical properties of galaxies.
\begin{table}
\begin{center}
\caption{Results from SED  fitting for the average  MIPS-LBG}\label{table_main_result}
\begin{tabular}{| c | c c  c|}
    \hline
     Models             & {MW $^a$}& {MIPS} $^b$ & other works\\[5pt]
     \hline
    IR luminosity: log( L$_{IR} /L_{\odot}$)   & $12.13$  &12.11 & $12.21 ^{c}$\\[5pt]
                                                              &     &                                  & $12.44^{d}$\\
                   \hline
    Stellar mass: $M_{\ast}$  ($M_{\odot}$)       &$8\times 10^{10} $           &  $8\times 10^{10} $   &$7.9 \times 10^{10}$ $^{e}$ \\[5pt]
    \hline
    age (Gyr) & 0.5& 0.5 & 1  $^f$ \\[5pt]
               \hline
     SED-derived SFR ($M_{\odot}/yr$)         &$200$           & 200   &  $-$ \\[5pt]
       IR-derived SFR                                       &        $233$    &  222   & 275$^g$\\
           \hline
    Dust mass: $M_{d}$  ($M_{\odot}$)      &$7\times 10^{7}$           &  $7\times 10^{7}$ & $5.5\times 10^{8}$  $^{h}$   \\[5pt]
    \hline
    Observational frame 850 $\mu m$ flux  density : $f_{850} (mJy)$     &$0.16$           &  $0.34$   &$1.36^{i}$  \\[5pt]
    \hline
   UV Slope: $\beta$           & -1.6      &-2.1 &$-$  \\[5pt]
    \hline
   Extinction: $E(B-V)$  & 0.213           & 0.127   &0.1  $^{j}$     \\[5pt]
    \hline
   Luminosity weighted metallicity:  $<Z>_{lum}$  &$0.017 $  & $0.017$      &$-$   \\[5pt]
    \hline
     ISM metallicity:  $Z$  &$0.02$  & $0.02$    &0.02 $^{k}$     \\[5pt]
        \hline
    Dust to gas ratio: $D_g$        &$0.0025$           &  $0.0025$   &$-$  \\[5pt]

\hline
\end{tabular}
\\
$a$: $10^{11} M_{\odot}$ model (M11) at 0.5 Gyr with WM-like parameters (see Table \ref{table_mc} and \ref{table_grasil_dp}) \\
$b$:  $10^{11} M_{\odot}$ model (M11) at 0.5 Gyr with MIPS parameters  (see Table \ref{table_mc} and \ref{table_grasil_dp})\\
$c$, $g$  and $i$ from \cite{Magdis2010ApJ...720L.185M}, \\
$d$ and $h$ from  \cite{Rigopoulou2010MNRAS.tmpL.154R}\\
$e$, $f$ and $k$  from \cite{Magdis2010MNRAS.401.1521M}\\
$j$  from Magdis's  private communication \\

\end{center}
\end{table}
\begin{table}
\begin{center}
\caption{Adopted values of MC in GRASIL}\label{table_mc}
\begin{tabular}{| c | c c |}

    \hline
                  & {MW}& {MIPS}  \\
    \hline
    $t_{esc}$ $^1$ & 2             & 200        \\[5pt]
    $f_{mc}$ $^2$  & 0.5           & 0.5             \\[5pt]
    $M_{mc}$ $^3$ &$10^6 $  & $10^6$         \\[5pt]
    $R_{mc}$ $^{4}$       &40           &  16     \\[5pt]
\hline
\end{tabular}
\\
$^1$ Timescale for the evaporation of MCs, in Myrs.\\
$^2$ Fraction of  gas content in the MCs.\\
$^3$ Total gas mass in each MC, in $M_{\odot}$.\\
$^{4}$ Radius of each MC, in pc. The value of $R_{mc}$ in the MW is just to respond to the lower dust-to-gas ratio in the MW than in the star-burst galaxies (see text in Sect. \ref{dust_enviroment})
\\
\end{center}
\end{table}
\begin{table}
\caption{Parameters for size distribution of the dust components (Eq. \ref{pah} and \ref{eq:a1}) in
the MW and MIPS-LBGs. The MW size distribution is
that derived by \cite{Silva1998ApJ...509..103S} to match observations from the MW. The MIPS-LBGs
 size distributions are calculated in this paper to
match SED.}\label{table_grasil_dp}
\begin{tabular}{| c | c c | }
    \hline
                    & MW  & MIPS  LBG \\
    \hline
    PAH            &       &                 \\
    X                & $3.3\times10^{-25}cm^{2.5}/H$ &$5\times10^{-25}cm^{2.5}/H$ \\
    \hline
    $\mathbf{Graphite}$   &         &  \\[5pt]
    $a_{min}$ $(\AA)$      & 8       & ... \\[5pt]
    $a_{b}$ $(\AA)$        & 50      & 8000  \\[5pt]
    $a_{max}$ $(\AA)$      & 2500    & 22500     \\[5pt]
    $\beta_{1}$        & -3.5    & -3.5     \\[5pt]
    $\beta_{2}$        & -4.0    & ...        \\[5pt]
    $\mathbf{Silicate}$ &         &             \\[5pt]
    $a_{min}$ $(\AA)$      & ...     & ...              \\[5pt]
    $a_{b}$ $(\AA)$        & 50      & 800          \\[5pt]
    $a_{max}$ $(\AA)$      & 2500    & 12500          \\[5pt]
    $\beta_{1}$        & -3.5    & -3.5        \\[5pt]
    $\beta_{2}$         & ...     &  ...         \\[5pt]
\hline
\end{tabular}\\
\end{table}

\subsection{The fitting approach} \label{fitting}
In this work, with the help of the SFH, chemical pattern and dust properties from the chemical evolution models \citep{Pipino2011A&A...525A..61P} and pan-spectral  energy distribution data of the stacked MIPS-LBG  \citep{Magdis2010ApJ...720L.185M},  we break the SFR-age-metallicity degeneracies for SED modeling. 
Basically, the flux derived by median stacking analysis represents the average properties of many undetected individual objects. We will not give ``ad hoc" parameters but the general properties of the average MIPS-LBG by modelling the stacked SED  without taking into account the filter widths in SED fitting.  We will adopt an ``educated fitting" approach to estimate the physical parameter of the galaxy. This approach is guided by  answering  these questions: 1) Which parameter dominates the overall-level SED of a galaxy? In other words, from a galaxy SED, which estimated property is most reliable?  2) How do other parameters affect the SED in detail, such as the shape and the peak of some parts of a SED?  By means of this approach, we use  SFHs suggested by self-consistent chemical evolution models.  Moreover, chemical and dust properties, such as metallicity and dust mass, which are the basic ingredients in SED modeling,  are given by the  same chemical evolution model which gives the SFH.   The ``fitting" approach  is the following: first, since the mass is the most robust parameter in SED fitting \citep[e.g. see][]{Shapley2005ApJ...626..698S}, the total  mass of the average MIPS-LBG is estimated by comparing the predicted overall-level of SED with data, and confirmed by further fitting steps. This treatment effectively reduces the computation time,
since we do not need to test the MC and  dust parameters in GRASIL to estimate the total mass ( Step One, see Fig \ref{mass}).  Once the galaxy is selected by total mass estimate, the SFH, chemical and dust evolution history are given by the chemical evolution model with that total mass.  Second, the age of the average MIPS-LBG  is estimated  by comparing the total mass-selected  overall-level of SED with data.   At this step, the SFR, metallicity, stellar mass, dust mass of the average MIPS-LBG  are derived by the SFH adopted in the  chemical evolution model ( Step Two, see Fig \ref{age}).   Third, the ``best-fitting" dust parameters (shown in Table \ref{table_mc} and \ref{table_grasil_dp})   are derived by min-$\chi^2$ method based on a grid of testing parameters in GRASIL (see an example of grid models in Fig. \ref{648p}).  At this step, the galactic parameters estimated in previous steps are confirmed by the ``best-fitting" SED ( Step Three, see Fig \ref{best0.5}).

To fit the SED, we have tested in GRASIL many SFHs originating from  elliptical models. These models cover  a large mass range  \citep[from $10^{9} M_{\odot}$ to $10^{12}  M_{\odot}$, see][for detail]{Pipino2011A&A...525A..61P}. The parameters and results of  three examples of tested  models  with total mass  $3\times 10^{10} M_{\odot}$ (M310), $10^{11}  M_{\odot}$ (M11) and $5\times 10^{11}  M_{\odot}$ (M511) are shown in Table \ref{chem_p}.
SFHs  of these three models and the estimated SFR of  the  average MIPS-LBG are shown in Fig. \ref{sed_sfr}.  Dust mass, dust-to-gas ratio and metallicity as a function of galactic age  of a typical elliptical  of $10^{11} {M_{\odot}}$
(the ``best-fitting" galaxy, see below) adopted in the GRASIL are shown in  Fig. \ref{che-inform1}-\ref{che-inform3}, respectively.

The free parameters in GRASIL, which will be investigated by min-$\chi^2$ method in this work,  are the ones related to MCs and dust size  distribution. There are four parameters (in Table \ref{table_mc}) related to  MCs: 1) the time scale $t_{esc}$ of young stars escaping from their birth places (MCs), 2) the fraction of gas  content in  MCs $f_{mc}$, 3) the mass of a single MC $M_{mc}$, 4) the radius of a single MC $R_{mc}$. Note that, the predicted SED depends on $M_{mc}$ and $R_{mc}$ only through the combination $M_{mc}/R_{mc}^{2}$, which is the true free parameter.   The dust size  distribution for each dust component is described by the power laws (Eq. \ref{pah} and Eq. \ref{eq:a1} in Sections 3.2 and 3.3). The parameters of graphite and silicate  (in Table \ref{table_grasil_dp}) are the size lower limit ($a_{min}$), size upper limit ($a_{max}$), the connection point of the two power laws ($a_{b}$) and the ones related to slopes ($\beta_{1}$ and $\beta_{2}$ in Eq. \ref{eq:a2}). The parameter $X$ of PAH
size distribution controls the abundance of PAH. The normalization parameters, which control  the dust abundance and dust component relative abundances are given by the chemical evolution model (see Sect. \ref{dust_enviroment} and \ref{dust} for detail). Other  parameters are the same as in \cite{Silva1998ApJ...509..103S}. The standard (Milky Way, MW for short) and our ``best-fitting" model parameters adopted in SED modeling are shown in Tables \ref{table_mc} and \ref{table_grasil_dp}. Effects caused by various IMFs, simple stellar populations and dust optical properties will not be discussed in this paper.

\section{Results and discussion}\label{result}

The zero order estimates of the relevant physical quantities (masses, SFRs) are  typically derived by global or integral properties, such as overall-level SED and  IR luminosity,   therefore  estimates of those quantities  are robust. Other  quantities (such as UV slope and the predicted flux at a particular band)  depend on the shape of SED, and could be different. Table \ref{table_main_result} shows the main fitting properties of the average MIPS-LBG by different models.  These models fit the overall-level SED, and predict consistent IR luminosities.  Stellar mass and IR-derived  SFR in  all models are comparable, while other quantities are not necessarily consistent with each other.  We will discuss these properties in more detail below.

\subsection{Stellar populations}

The overall-level rest-frame SED, like the value of the luminosity density or flux density, of a galaxy is dominated by its total galaxy mass. Particularly,  the total stellar mass dominate the UV-optical part of the SED and total dust mass dominate the IR part of  the SED. In general, stars, such as AGB stars and SNe, are  the main sources of dust in a galaxy. A more massive star-forming galaxy contains more stars as well as more dust \citep[e.g. see the case of young elliptical galaxies in][]{Pipino2011A&A...525A..61P}. The more massive star forming galaxy will show higher overall-level rest-frame SED. Since our estimations are based on the UV-radio overall-level SED fitting, our results of the stellar population  are derived by a complementary approach comparing to the normal approach, which is usually based on optical to near-IR SED fitting.

\begin{figure}
\begin{center}
\includegraphics[width=3in]{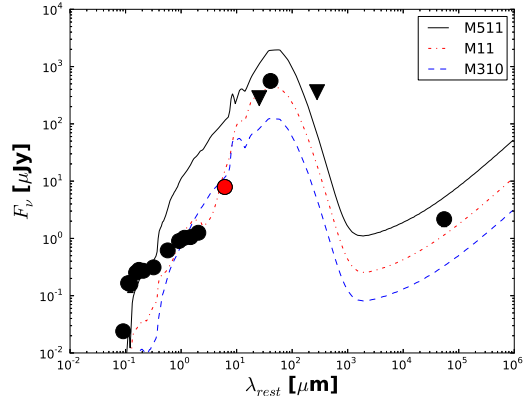}\\
\caption{Rest-frame SEDs of different models.  M511, M11 and M310 represent $5\times 10^{11}, 10^{11}$ and $3\times 10^{10}M_{\odot}$  galaxies with star burst SFH at 0.5 Gyr with the MW dust parameters shown in Table \ref{table_mc} and Table \ref{table_grasil_dp}, respectively.  Data are from \cite{Magdis2010ApJ...720L.185M}: for the SED we use the median UGR+BViJK+IRAC+MIPS24 photometry of MIPS-LBGs, and the values (or upper limits) derived from stacking 100 $\mu$m, 160 $\mu$m AzTEC and radio, divided by 1 + z.  Downward triangles are up-limits. The MIPS data is indicated with  a red ponit. At the first fitting step, the $10^{11} M_{\odot}$ model  is empirically selected  for next fitting step.}\label{mass}
\end{center}
\end{figure}

\begin{figure}
\begin{center}
\includegraphics[width=3in]{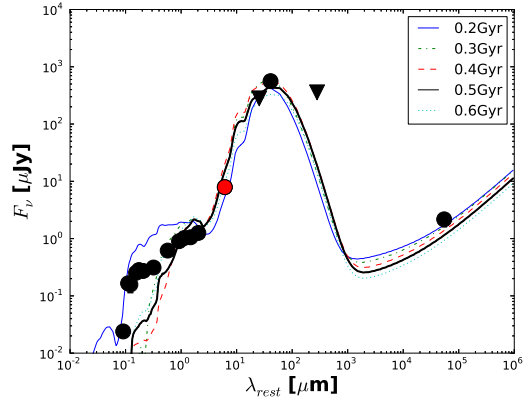}\\
\caption{Rest-frame SED of a $10^{11}M_{\odot}$ galaxy at different ages with the  MW dust parameters as in Table \ref{table_mc} and Table \ref{table_grasil_dp}. The parameter space of age is  $< 0.7$ Gyr.  The model with age $ 0.3-0.6$ Gyr could fit the overall-level SED. At the second fitting step, the $10^{11} M_{\odot}$ galaxy at 0.5 Gyr is the  model adopted for further fitting step.wo This  model predicts comparable SFR and stellar mass with other people's work (see text for detail). MIPS-LBGs data are from \cite{Magdis2010ApJ...720L.185M} as described in Fig \ref{mass}.}\label{age}
\end{center}
\end{figure}

 We compare the predicted SEDs of different massive galaxies at  different ages (see an example of different massive galaxies at moderate age 0.5 Gyr in Fig. \ref{mass}.  Since short and intense star formation histories are yielded by chemical evolution models, the parameter space of age is not large (age $<0.7 Gyr$). From Fig. \ref{mass} and Fig. \ref{age}, we can estimate that the galaxy with total mass $\sim 10^{11} M_{\odot}$  could reproduce the overall-level SED of the average  MIPS-LBG. Note that, this mass is the total  mass of an elliptical galaxy at the wind time given by the chemical evolution model. In fact, after the wind, no more star are formed and the galaxy evolves passively.

\begin{figure}
\begin{center}
\includegraphics[width=3in]{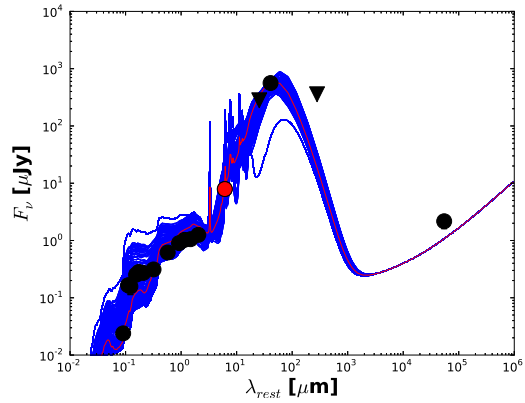}\\
\caption{Rest-frame SEDs of a $10^{11}M_{\odot}$ galaxy  at 0.5 Gyr with various dust environments  and dust intrinsic properties. Blue lines: various dust size distributions and MC properties.  Only with these various  properties, the UV-optical part of the SED varies by a factor of ten. It is clear that the  MW properties cannot fit the SED. Red line: MW dust and MC properties in Table \ref{table_mc} and Table \ref{table_grasil_dp}. MIPS-LBGs data are from \cite{Magdis2010ApJ...720L.185M} as described in Fig \ref{mass}.}\label{648p}
\end{center}
\end{figure}

 From the modeling point of view, the total stellar mass is dominated by SFH, IMF and the age of stellar populations. Age is the only free parameter of stellar populations when SFH, chemical enrichment history and IMF of a galaxy are fixed. With a given SFH in a chemical model, the SFR is a predictable parameter.  By comparing  with the value eatimated from observational indicators, SFR could constrain  the age of a galaxy. UV and IR indicators for SFR \citep[e.g.][]{Kennicutt1998ARA&A..36..189K} have been adopted in the literature for high redshift galaxies.  Besides,  depending on  stellar population synthesis models, the UV-derived SFR depends on the treatment of extinction for deriving  the extinction-correct fluxes, and the IR-derived SFR depends on  the SED library for converting observed flux (e.g. the flux at 24 $\mu m$) to total ($8-1000 \mu m$ ) IR luminosity  $L_{IR}$. This approach has the drawback of assuming the low-redshift template SEDs accurately represent the SED of high-redshift galaxies. For example,  the excessive high 850 $\mu m$ fluxes  predicted by an empirical UV-based relationship  and the unreasonable high $L_{IR}$  estimated by  MIR flux-IR luminosity  correlation in SED templates have  caused   the  problem  of non-detection  of  $z \sim 3$ LBGs in  850 $\mu m$ band,  reported in  \cite{Chapman2000MNRAS.319..318C}  and \cite{Shim2007ApJ...669..749S} respectively. Additionally,  various dust intrinsic properties strongly affect the UV-FIR fluxes (see Fig. \ref{648p}).  Therefore, we need other indicators to estimate the SFR in our SED fitting approach. In GRASIL the non-thermal radio emission, which is not affected by  dust intrinsic properties, is proportional to the SN II rate.  Since the life of SNe II is very short ($<30$ Myr),  the SN II rate is proportional to the SFR. So the rest-frame radio
flux  should be a good estimator of SFR. This is reflected by the fact that   both the radio flux  and SFR decrease (shown in Fig. \ref{sed_sfr} and \ref{age}) at the age from 0.2 to 0.6 Gyr.   For the $10^{11}M_{\odot}$ model, the  phase with SFR $\sim 200 M_{\odot}/yr$,  corresponding to the age of $0.3$ to $0.6$ Gyr (see Fig \ref{sed_sfr}), fits the rest-frame radio flux well, and this SFR agrees  with the estimation ($\sim 250 M_{\odot}/yr$) of \cite{Magdis2010ApJ...720L.185M}.     Since we have already selected the $10^{11} M_{\odot}$ model based on overall-level SED,  the SFR derived here benefits  from the merits of   radio-derived  and  SED-derived approaches, and is  remarkably  consistent with the IR-derived SFR (see Table \ref{table_main_result} and discussion below).

 In our approach, the  age is estimated  by  fitting the overall-level SED and the radio flux, namely it is a luminosity weighted age confirmed by ``observed'' SFR. The SFR and total mass suggest that the ages of MIPS-LBGs should be $ \sim 0.3-0.6$ Gyr (see Fig. \ref{age}), and  younger than 0.7 Gyr since the LBGs are star-forming galaxies.   Note that, the estimated SFR $=200 M_{\odot}/yr$  is  a typical SFR in the age range of $0.3$ to $0.6$ Gyr. The SFR is $221 M_{\odot}/yr$ and $140 M_{\odot}/yr$  at $0.3$ and $0.6$ Gyr for the $10^{11}M_{\odot}$ model, respectively. The total stellar mass is $\sim 6\times 10^{10} M_{\odot}$  and $\sim 1\times 10^{11} M_{\odot}$  at $0.3$ and $0.6$ Gyr for the $10^{11}M_{\odot}$ model, respectively.  We adopt 0.5 Gyr for the next fitting step.  This age corresponds to SFR $\sim 185 M_{\odot}/yr$ and total stellar mass $\sim 8\times 10^{10} M_{\odot}$, which agrees with the value ($\sim 7.9\times 10^{10} M_{\odot}$) of \cite{Magdis2010ApJ...720L.185M}. Our age estimate $0.3-0.6 Gyr$ is not consistent with the  median age $\sim 1 Gyr$  derived by synthesis population model in \cite{Magdis2010MNRAS.401.1521M}. In particular, our elliptical model does not allow a LBG to be  older than $1 Gyr$.
  It is however worth reminding, that while the star formation history predicted in chemical evolution models is modulated by the gas infall and
self-consistently quenched due to SN-driven wind,
this does not happen in the template star formation histories adopted in stardard SED fitting, and this might cause the disagreement in the age estimate.

\subsection{Dust properties}\label{dust_enviroment}

\begin{figure}
\begin{center}
\includegraphics[width=3in]{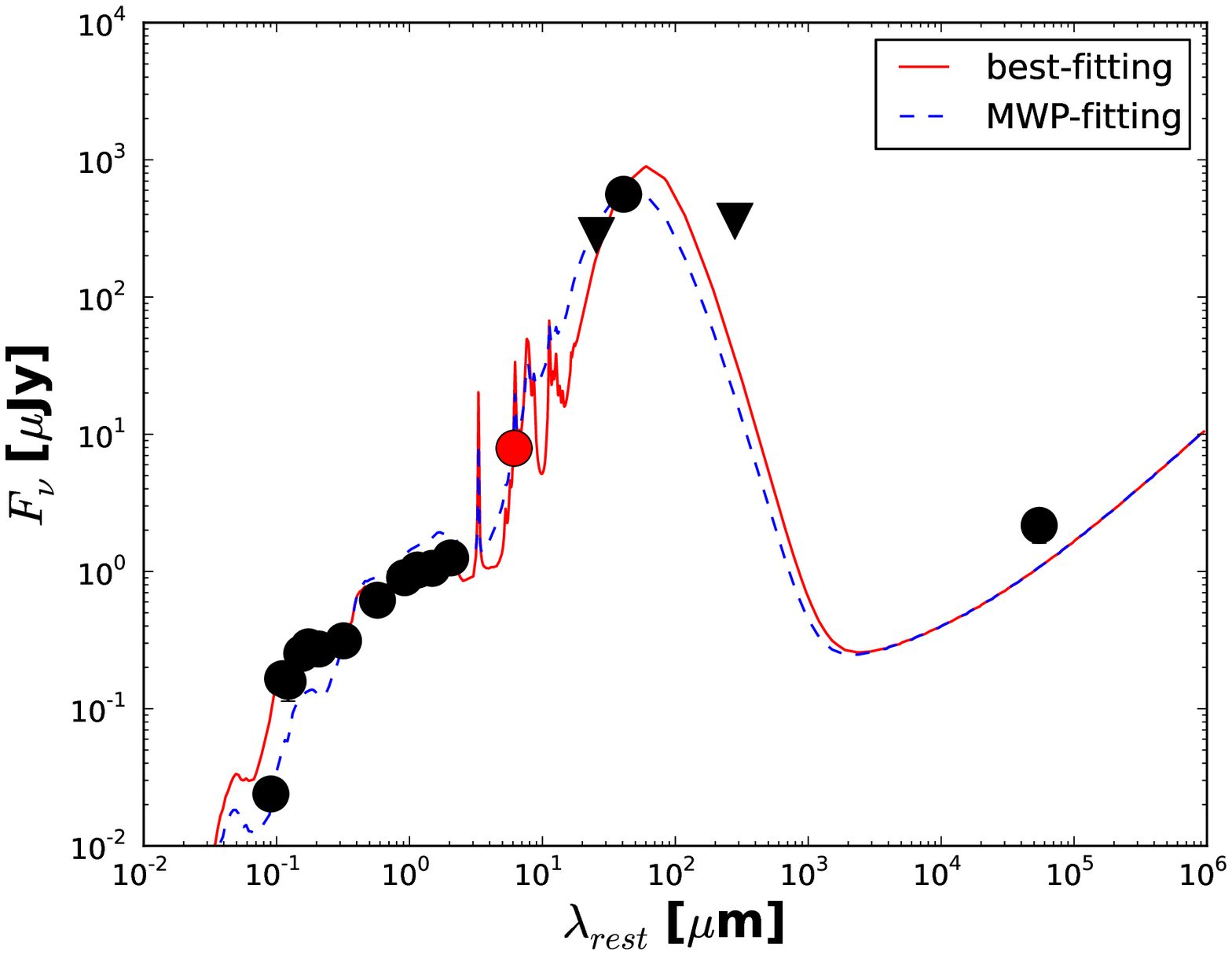}\\
\caption{``Best-fitting" rest-frame SED.  Red solid line: the ``best-fitting" SED of a $10^{11}M_{\odot}$ galaxy at 0.5 Gyr with MIPS-LBGs dust and MC properties shown in  Table \ref{table_mc} and Table \ref{table_grasil_dp}.   At the third fitting step, this ``best-fitting" model is given by min-$\chi^2$ fitting approach based on the grid models shown in Fig. \ref{648p}. The values in Table \ref{table_mc} and Table \ref{table_grasil_dp} just give the general properties of the dust and MCs and should not be considered as the real ones. Note that, one can get an another ``best-fitting" model using another galaxy model (such as a $10^{11}M_{\odot}$ galaxy at 0.4 Gyr). However, the  overall-level SEDs are similar for models within a small age parameter space (see Fig. \ref{age}). Therefore, the trend of the dust and MC properties is the same for those models with different age. However, the parameters of Milky Way can not fit the SED no matter what the galactic age is adopted (see examples in Fig. \ref{age}). The mean differences between the ``best-fitting" and ``Milky Way parameters-fitting" SED (Blue dashed line, ``WMP-fitting") are the UV-optical shape and the position of the FIR peak. MIPS-LBGs data are from \cite{Magdis2010ApJ...720L.185M} as described in Fig \ref{mass}.}\label{best0.5}
\end{center}
\end{figure}
As  molecular clouds (MCs)  are the birth place of stars, studying properties of MCs at high redshift is important for  our understanding of the star forming in galaxies \cite[see review by][]{Riechers2011arXiv1103.3897R}. In this paper, we assume that  all the MCs have the same mass $M_{mc}$ and spherical radius $R_{mc}$. MCs make much more contributions in extinction than diffuse ISM  in a star-burst galaxy with the fractionary gas content in the MCs  $f_{MC}\geq0.5$.
The parameter $t_{esc}$ shown in Table \ref{table_mc} controls  how long the young stars remain in their birth clouds and roughly controls the fraction of extincted stellar light, therefore it controls the total level of rest-frame UV-optical SED and the slope of the rest-frame UV-optical SED.
The optical depth of dust in a MC is  determined by the dust-to-gas ratio $\delta$ times $M_{mc}/R_{mc}^2$. This parameter moderately controls the total level of rest-frame UV-optical SED and also affects the slope of rest-frame UV-optical SED. The ``best-fitting" parameters of MCs for  the  average  MIPS-LBGs are shown in Table \ref{table_grasil_dp}. The larger value of  the $t_{esc}$ and the smaller value of the $R_{mc}$ compared with the MW ones\footnote{The values of MC parameters of the MW are the same as in  spiral models in \cite{Schurer2009MNRAS.394.2001S}, except for $R_{mc}$. The dust-to-gas ratio is given by the chemical evolution models and  not modified by GRASIL in this work.  Since we do not use the spiral SED models,  we adopt a larger value of $R_{MC}$ for the MW responding to the effect of lower optical depth caused by  lower dust-to-gas in the MW than in star-burst galaxies. In any case, the  $t_{esc}$ is the dominant parameter.} imply that the  MCs in the  average MIPS-LBG are likely to be
in  more dense dusty environments.  The same trend is found in local and high redshift starburst galaxies \citep{Silva1998ApJ...509..103S,Schurer2009MNRAS.394.2001S,Swinbank2011arXiv1110.2780S}. \cite{Yan2010ApJ...714..100Y} suggested that the average molecular gas mass of  24$\mu m$ detected  ultra-luminous infrared galaxies (ULIRGs, L$_{IR}>10^{12}L_{\odot}$) at $z=1.6-2.5$ is $1.7\times 10^{10}M_{\odot}$ by observed CO J=$2\rightarrow1$ or J=$3\rightharpoonup2$ emission.
 Our total gas mass $M_{g}$ in $10^{11}M_{\odot}$ galaxy  at 0.5 Gyr is $1.6\times 10^{10}M_{\odot}$. With fractionary gas content in the MCs $f_{MC}=0.5$, the total mass in molecular clouds in our ``best-fitting" model is $0.8\times10^{10}M_{\odot}$. However, the lack of emission lines makes it difficult to break the degeneracy  of  the MC conditions  and dust size distribution.

Once the  emission from stellar populations  is set, the peak in the emission in far-IR (FIR)  is dominated by the total cold dust mass. The total dust mass $M_d$ is $\sim 7\times10^{7}M_{\odot}$ in ``best-fitting"  $10^{11} M_{\odot}$ model at 0.5 Gyr.   It is about a factor of eight  less than the value \citep[$M_d = 5.5 \pm 1.6 \times 10^{8}M_{\odot}$ in][]{Rigopoulou2010MNRAS.tmpL.154R} based on  single temperature grey-body fitting \citep{Hildebrand1983QJRAS..24..267H}.  This may partly be caused by three facts: 1) the uncertainty in the galaxy mass estimate, therefore the dust mass value in this work. Since the dust mass is a consequence of the galaxy evolution in our chemical models, the estimated total galactic mass  affects the total dust mass. For example, a $2\times 10^{11} M_{\odot}$ galaxy  \footnote{deviation of estimated total galaxy mass from  the ``best-fitting" value  within a factor of two is normal and acceptable in all SED fitting work.} at 0.5 Gyr  produces a dust mass $M_d \sim 2\times 10^{8}M_{\odot}$.  2) The well known degeneracy of dust temperature $T_d$ and slope $\beta_d$ in single temperature grey-body fitting \citep{Blain2003MNRAS.338..733B}. A higher $\beta_d$ will result in lower dust temperature derived form  single temperature grey-body fitting \citep{Sajina2006MNRAS.369..939S}, therefore a  higher dust mass will be estimated.  3) The uncertainty of  single temperature grey-body fitting parameter, such as the rest-frame dust mass absorption coefficient $\kappa$ \citep[e.g. a factor $\sim$ 7 at 800 $\mu$ m  estimated by][]{Hughes1997MNRAS.289..766H}.  Note that, the single temperature grey-body fitting could predict a higher IR  flux than the value  by full SED fitting.   An example is shown in \cite{Magdis2010ApJ...720L.185M}.  The value of the far-IR emission peak predicted by  the single temperature grey-body fitting shown in their Fig. 4 is larger than 1 $mJy$, while the peak value shown in their Fig. 2 is smaller than 1 $mJy$ predicted by  SED fitting.

Prominent PAH features are observed in MIPS-LBGs \citep{Huang2007ApJ...660L..69H}. In star-burst galaxies, the mid-IR (MIR) flux is contributed by small hot dust and PAH \citep[see e.g.][and references
therein]{Laurent2000A&A...359..887L}.  The abundance of PAHs of our model is calculated from the chemical composition of
the dust as predicted by the chemical evolution model, and their abundance
is proportional to the total abundance of carbon molecules in the
dusty component of the ISM.  The PAH size distribution adopted in GRASIL is :
\be\label{pah}
\frac{dn}{da}=X a^{-3.5} cm^{-1}.
\ee
PAH is needed in the ``best-fitting"  model  to fit the  average MIPS-LBG  MIR flux (see Fig.\ref{best0.5}). However,  we cannot put more constraints on the PAH without emission lines.  The ``best-fitting" model gives $X=5\times10^{-25}cm^{2.5}/H$ ($X=3.3\times10^{-25}cm^{2.5}/H$ in \cite{Silva1998ApJ...509..103S}).

\subsection{Dust intrinsic properties}\label{dust}
For many years,  many efforts have been spent on investigating the dust size distribution through three approaches: i) theoretical approach \citep[e.g.][]{Birnstiel2011A&A...525A..11B}, ii) fitting extinction curve approach \citep[e.g.][]{Mathis1977ApJ...217..425M,Weingartner2001ApJ...548..296W,Clayton2003ApJ...588..871C,Zubko2004ApJS..152..211Z} and iii) by means of the SED approach \citep[e.g.][]{Carciofi2004ApJ...604..238C,Takeuchi2004A&A...426..425T,Piovan2006MNRAS.366..923P}. However, the nature of dust size distribution is still not clear, especially for in high redshift objects.

The position of the peak  in FIR is mainly affected by the total dust mass, dust component relative abundances and dust size distribution, but not by the stellar populations  \citep[see Fig.\ref{mass} and][]{Slater2011arXiv1103.2170S}. With the help of chemical evolution models, which predict  the total dust mass and dust component relative abundance, we can study the dust size distribution in a more reliable way.
The graphite and silicate grains dust size distribution strongly affect the slope of the rest-frame UV-optical SED.  For the sake of simplicity,  a simple power low dust size distribution form is adopted in this paper.  The dust size distribution for the $i$ (graphite or silicate) composition follows a broken power law defined with a threshold $a_{b}$:
\makeatletter
\let\@@@alph\@alph
\def\@alph#1{\ifcase#1\or \or $'$\or $''$\fi}\makeatother
\begin{subnumcases}
{\frac{dn_i}{da}=}
A_i n_H a^{\beta_1},  &if $a_b<a<a_{max} $, \label{eq:a1}\\
A_i n_H a_b^{\beta_1-\beta_2}a^{\beta_2}, &if $a_{min}<a<a_b $.\label{eq:a2}
\end{subnumcases}
\makeatletter\let\@alph\@@@alph\makeatother
 For the sake of simplicity, dust in MIPS-LBGs only use Eq. \ref{eq:a1} (see Table \ref{table_grasil_dp}).
The dust mass $m_i$ and dust-to-gas ratio $\delta$ are provided by the chemical evolution model:
\be\label{eq:m}
m_i=\int_{a_{min}}^{a_{max}}\frac{4\pi a^3 \rho_i}{3n_H}\frac{dn_i}{da}da,
\ee
\be\label{eq:d}
\delta=\frac{\sum_i m_i}{m_H},
\ee
where $n_H$, $m_H$ and $\rho_i$ are the gas number density, gas mass and grain mass density, respectively.
The normalized parameter $A_i$  is calculated by Eq. \ref{eq:a1}, \ref{eq:m}, \ref{eq:d}.
Comparing with the dust size distribution in the MW (see Table \ref{table_grasil_dp}), a larger amount of dust with larger size is needed to fit the slope of rest-frame UV-optical SED  (see Fig. \ref{648p} and Fig. \ref{best0.5}). With the theoretical indication of a few $\mu$m size \citep[e.g.][]{Tanaka2005ApJ...625..414T,Birnstiel2009A&A...503L...5B} and  observational suggestion of up to cm-sizes \citep[e.g.][]{Rodmann2006A&A...446..211R,Ricci2010A&A...512A..15R}, our estimation of maximum  dust size $a_{max}$=2.2 $\mu$m is acceptable.

With indications from  both the peak and the position of  FIR and the slope of rest-frame UV-optical SED, we suggest that the dust size distribution of the average MIPS-LBG is top-heavy (flatter) comparing with that in the MW.  The reliability of this trend for all high redshift LBGs  needs more data to provide more reliable constraints for the models in the future.

\subsection{The impact of the adopted SFH}\label{imp_sfh}

\begin{figure}
\begin{center}
\includegraphics[width=3in]{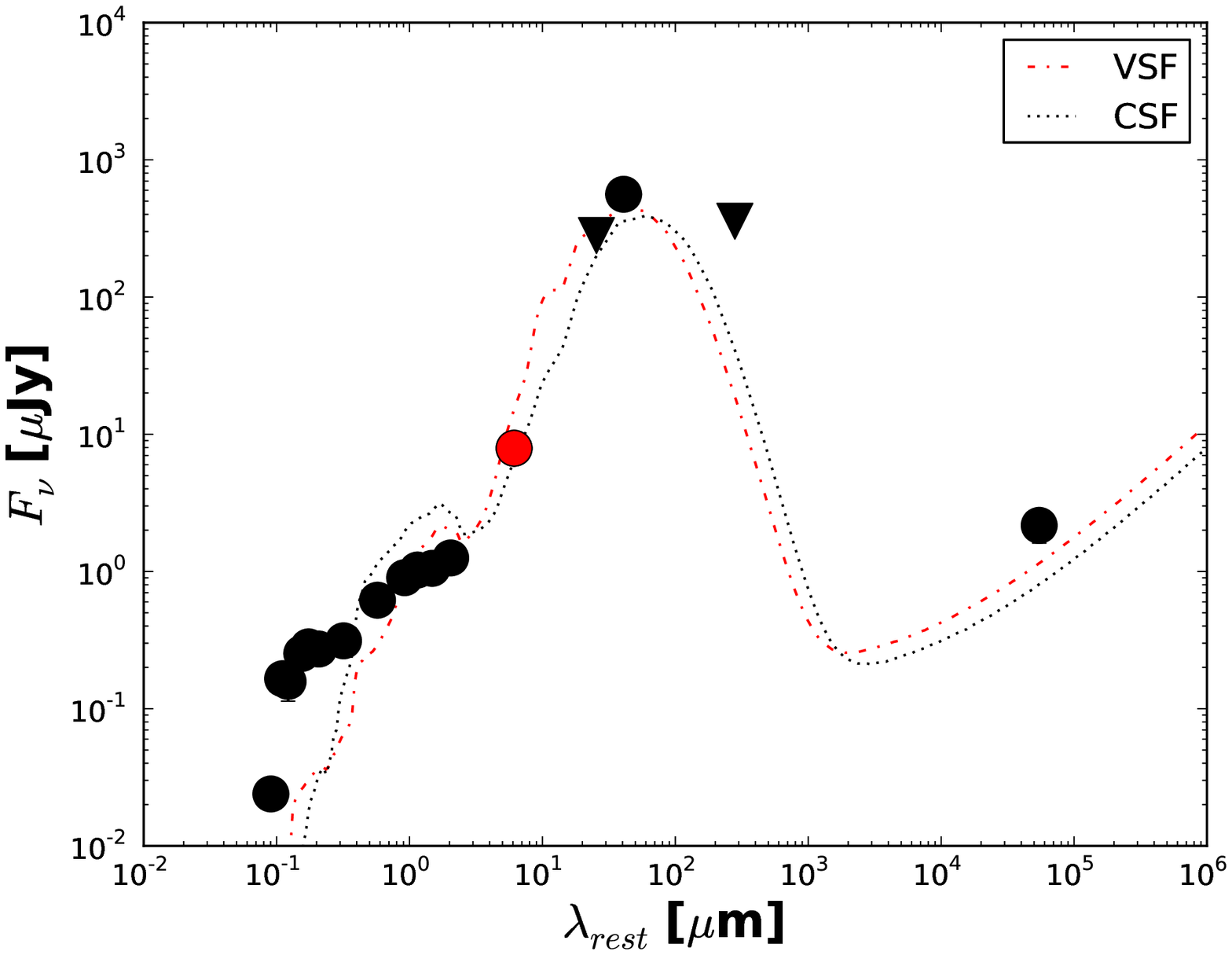}\\
\caption{Rest-frame SEDs of different models.  To show the necessary of more reliable SFHs, SEDs  of  a  model (age $\sim$ 0.5Gyr, stellar mass $\sim 8\times10^{10} M_{\odot}$), which takes a self-consistent SFH ( M11 case in Fig. \ref{sed_sfr}, SSF for short) and   a  model (age $\sim$ 2Gyr, stellar mass $\sim 9\times10^{10} M_{\odot}$) with a constant SFH (CSF) are shown.  MIPS-LBGs data are from \cite{Magdis2010ApJ...720L.185M} as described in Fig \ref{mass}.}\label{com_mass}
\end{center}
\end{figure}

The adopted SFH is the main input ingredient  which produces different predictions on detailed physical processes by means of population synthesis modeling. By the standard SED fitting, we cannot rule out other SFHs than the ones adopted.   For example, Fig. \ref{com_mass}. shows  two models adopting  a self-consistent star formation history (SFH of M11 case in Fig.\ref{sed_sfr}, SSF for short) and a constant star formation history (CSF). These two models can produce similar overall-level SEDs with fine-tuning of the ages  corresponding  to  comparable stellar masses.  In this paper, to improve the reliability of the adopted SFH for the z$\sim$ 3 MIPS-LBGs, we adopted  chemical model constrained SFHs, which have been successfully  reproduced the chemical data of several   z$\sim$ 3 LBGs.

The stellar mass derived by  Elliptical-type star-burst SFH in this work  is consistent with other works, although we used a different SED model and fitting approach to fit the SED. Again, this is because  the stellar mass is a more robust parameter than other parameters, such as age, which have been found in other SED fitting works \citep[e.g.][]{Papovich2001ApJ...559..620P,Shapley2005ApJ...626..698S,Magdis2010MNRAS.401.1521M}.
Recently, \cite{LoFaro2013ApJ...762..108L} found  discrepancies between their results and those based on optical-only SED fitting for the same objects. By fitting observed SEDs with their physical model, they found higher stellar masses
(by $\Delta M_{\ast} \sim$ 1.4 and 2.5 dex) resulting from higher extinction (the average total extinction in the rest-frame V-band $A_{V} \sim$ 0.81 and 1.14 ) for $z\sim 1$ and $z\sim 2$ dusty normal star-forming (e.g. SFR$\leq 10 M_{\odot} yr^{-1}$) galaxies, respectively.  With only one average $z\sim 3$ dusty star-forming galaxies in our paper, it is hard to test whether  there is also a same systematic discrepancy, and the discrepancy is mainly caused by extinction for $z\sim 3$ dusty star-burst (e.g. SFR$> 100 M_{\odot} yr^{-1}$) galaxies or not. We would like point out that the dusty star-forming galaxies at different redshifts are not necessarily linked from an evolutionary point of view, therefore the types of SFH  and the evolution phases  in which galaxies are observed might be different. Therefore, one might get comparable stellar mass by the full SED fitting and optical-only SED fitting with different ages and extinction values (laws).   For example, by playing with dust parameters in the CSF case shown in Fig. \ref{com_mass}, we would expect to fit the optical-only SED at 2 Gyr estimating a stellar mass $\sim 9\times 10^{10}$ which is comparable with the stellar mass $\sim 8\times 10^{10}$ at 0.5 Gyr in  the ``best-fitting" model by  full-SED fitting.

Unlike the stellar mass, our estimated age ($\sim$ 0.5 Gyr) is younger than other works ($\sim$ 1 Gyr). Our intense  star-burst SFH allows the galaxy to form enough stars and dust to produce the overall-level SED at that age. Dust makes  the key link between the UV-optical SED and IR SED. In the standard SED fitting approach, dust mass cannot be derived, therefor it does not depend on the adopted SFH. In our approach, dust mass  is a consequence of galaxy and dust evolution in the chemical models and depends on the SFH. By assuming reasonable detailed dust properties, the dust in our model interacting with stellar light,  reproduces  the SED of the average MIPS-LBG. The value of the  dust mass in our work is about a factor of eight  less than the value in \cite{Rigopoulou2010MNRAS.tmpL.154R} based on  single temperature grey-body fitting (see discussion in Sect. \ref{dust_enviroment}). Note that, our predictions on dust environment and dust  intrinsic properties  correspond to  ``best-fitting" SED  of the average MIPS-LBG, and further work/data about these MIPS-LBGs are needed to confirm this result.

Clearly, the  high SFR  shown in the LBGs could be triggered by galaxy-merger, which means complectly different SFH and physical processes. This  scenario of galaxy evolution is not considered in this work. We hope that more  data in the future, such as the galaxy morphology and chemical abundances, will reveal the nature of all  LBGs.

Having discussed the  limitations in our approach,  we will compare our ``best-fitting" model for the average MIPS-LBG  with other objects which show similar  observed features in some aspects.

\subsection{Comparison with lower mass z$\sim$3 LBGs}
The gravitational lensed LBG cB58 is the best-studied typical LBG. This object
is at $z \sim 2.73$ and has a luminous mass
$\sim 10^{10}M_{\odot}$, a SFR $\sim 40 \,\rm M_{\odot}yr^{-1}$ \citep{Pettini2002ApJ...569..742P} and an effective radius of
$r_{L}\sim 2 \rm \, kpc$ \citep{Seitz1998MNRAS.298..945S},
for a $\Omega_m= 0.3,\Omega_{\Lambda}=0.7, h=0.70$
cosmology.
The average  MIPS-LBG, with total mass $\sim 10^{11} M_{\odot}$  estimated in this work, is  more massive than cB58  \citep[cB58 is $10^{10}-3\times 10^{10} M_{\odot}$ in][]{Pipino2011A&A...525A..61P}.
The model suggests a SFR $\sim 200 M_{\odot}/yr$  for the  average  MIPS-LBG, which is higher than that of cB58.
The metallicity of MIPS-LBGs ($\sim Z_{\odot}$) is higher than cB58 ($\sim 0.25  Z_{\odot}$).
These differences imply that  MIPS-LBGs and  cB58 are at the opposite ends of the observed mass-SFR \citep{Magdis2010MNRAS.401.1521M} and mass-metallicity (Maiolino et al., 2008) relations at $z\sim3$.

\begin{figure}
\begin{center}
\includegraphics[width=3in]{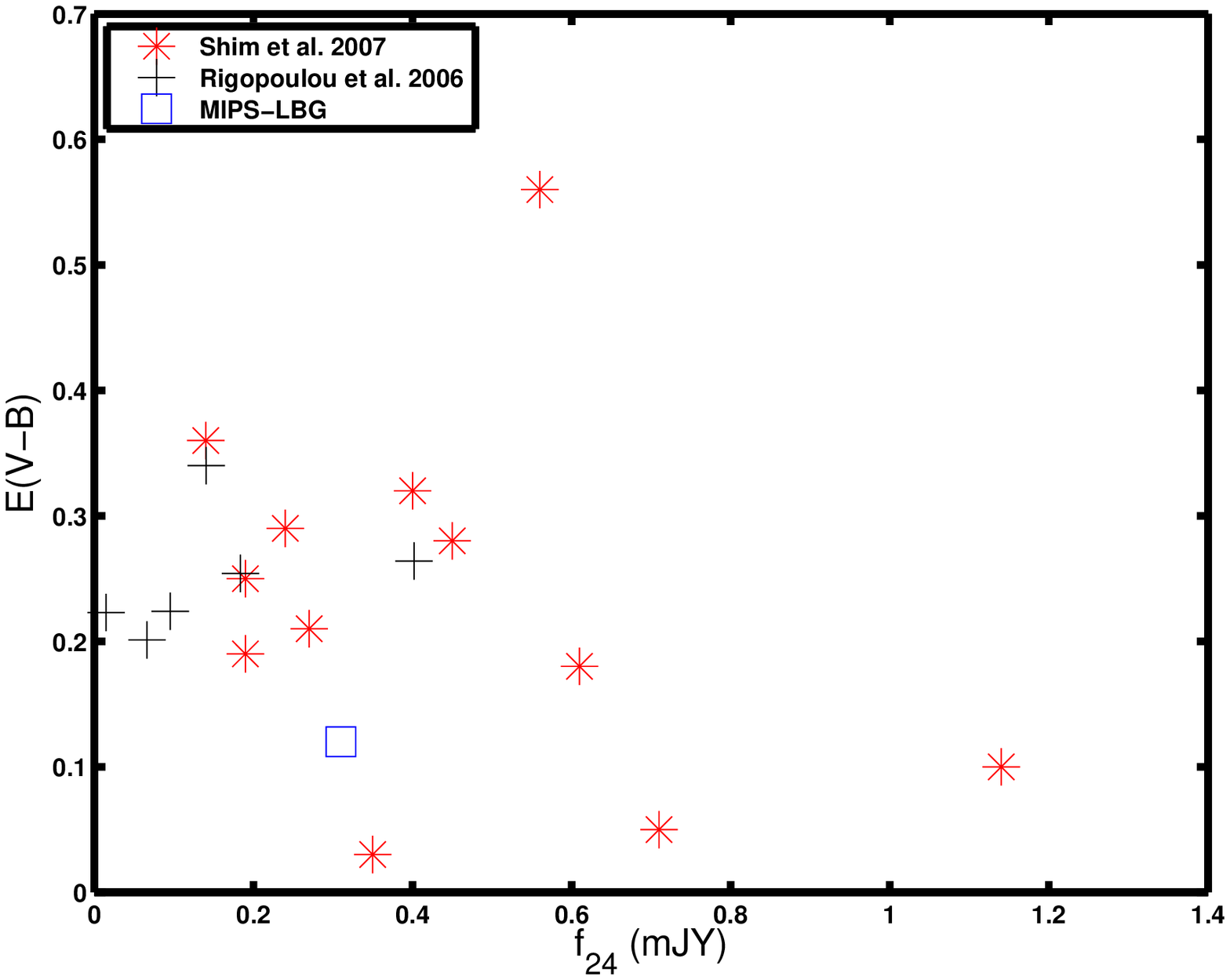}\\
\caption{Observed 24 $\mu m$ flux density $f_{24}$ vs estimated extinction E(B-V) for $24 \mu m$ MIPS detected LBGs at $z \sim 3$. Asterisk, cross, square symbols  represnet data from \cite{Shim2007ApJ...669..749S}, \cite{Rigopoulou2006ApJ...648...81R} and the average MIPS-LBG, respectively. The $f_{24}$ is not correlated with  E(B-V) for these objects. }\label{F24_EBV}
\end{center}
\end{figure}

The shape of a galaxy SED depends on many physical processes  as we have shown above. We compare the average MIPS-LBG in this work with other  $24 \mu m$ MIPS detected $z \sim 3$ LBGs  \citep{Rigopoulou2006ApJ...648...81R,Shim2007ApJ...669..749S} in Fig. \ref{F24_EBV} by  plotting  the estimated extinction  $E(V-B)$ versus observed $24 \mu m$ flux  density $f_{24}$. As shown in Fig. \ref{F24_EBV}, the observed frame $f_{24}$ does not correlate  with the $E(B-V)$ for these samples. One could not derive reliable IR properties only based on fitted parameters, which describe UV-optical properties. Therefore, we suggest that self-consistent SED models  are needed when
deriving  detail properties of galaxies  from  their UV-radio multi-band data 

\subsection{Comparison with  submillimetere galaxies (SMGs) }

The MIPS-LBGs with  IR luminosity of $L_{IR}= 1.3  \times 10^{12} L_{\odot}$  is in the class of   ultra-luminous infrared galaxies (ULIRGs, L$_{IR}>10^{12}L_{\odot}$). Since MIPS-LBGs could be the most massive, dusty, star-forming and young ellipticals as discussed in the above section,  we suggest that these ellipticals contribute a significant population for ULIRGs at redshift $z \sim 3$. 
 The non  detection of  infrared luminous LBGs in submm band is a puzzle, given their large $L_{IR}$ and  high UV-derived SFR, as well as  substantial dust component (see detail in \cite{Chapman2000MNRAS.319..318C} and \cite{Shim2007ApJ...669..749S}).
There are only two detections of FIR counterparts of MIPS-LBGs reported in the literature \citep{Chapman2000MNRAS.319..318C,Chapman2009MNRAS.398.1615C}. The higher dust temperature, different dust spatial distribution and  lower SFR in LBGs  than in SMGs have been suggested to explain this challenge \citep[e.g.][]{Chapman2000MNRAS.319..318C,Rigopoulou2006ApJ...648...81R,Magdis2010ApJ...720L.185M}.
Based on our ``best-fitting" model,  our prediction of observational frame flux density of MIPS-LBGs  at $850 \mu m$ is $0.3 mJy$, which is under the  confusion/detection limit  of  current  submillimetere (submm) surveys ($f_{850}=1 mJy$  and $f_{850}=2 mJy$).
Note that, our  $850 \mu m$ flux density $f_{850}=0.3 mJy$ is lower than $f_{850}=1.36 mJy$   \citep{Magdis2010ApJ...720L.185M} predicted by the ``best-fitting" SED from templates \citep{Chary2001ApJ...556..562C}, while the IR luminosity of our model and other models are comparable (see Table \ref{table_main_result}).  This reflects the fact that the predicted  flux could be quite different at a particular wavelength,  although a few observed fluxes or the overall-level  SED (e.g. the IR luminosity) could be comparable, since the  shape of SEDs could be quiet different.  The detailed SED of a galaxy depends on all physical processes in our model. Besides the explanations  mentioned in  the literature, we suggest the different galaxy morphology, dust environment and dust  intrinsic properties could also cause the non detection of LBGs in submm blank surrey.  However, it is hard to  make a conclusion about the link between LBGs and SMGs before more data, such as a large population with detailed SEDs and chemical abundances, are available. 

\begin{figure}
\begin{center}
\includegraphics[width=3in]{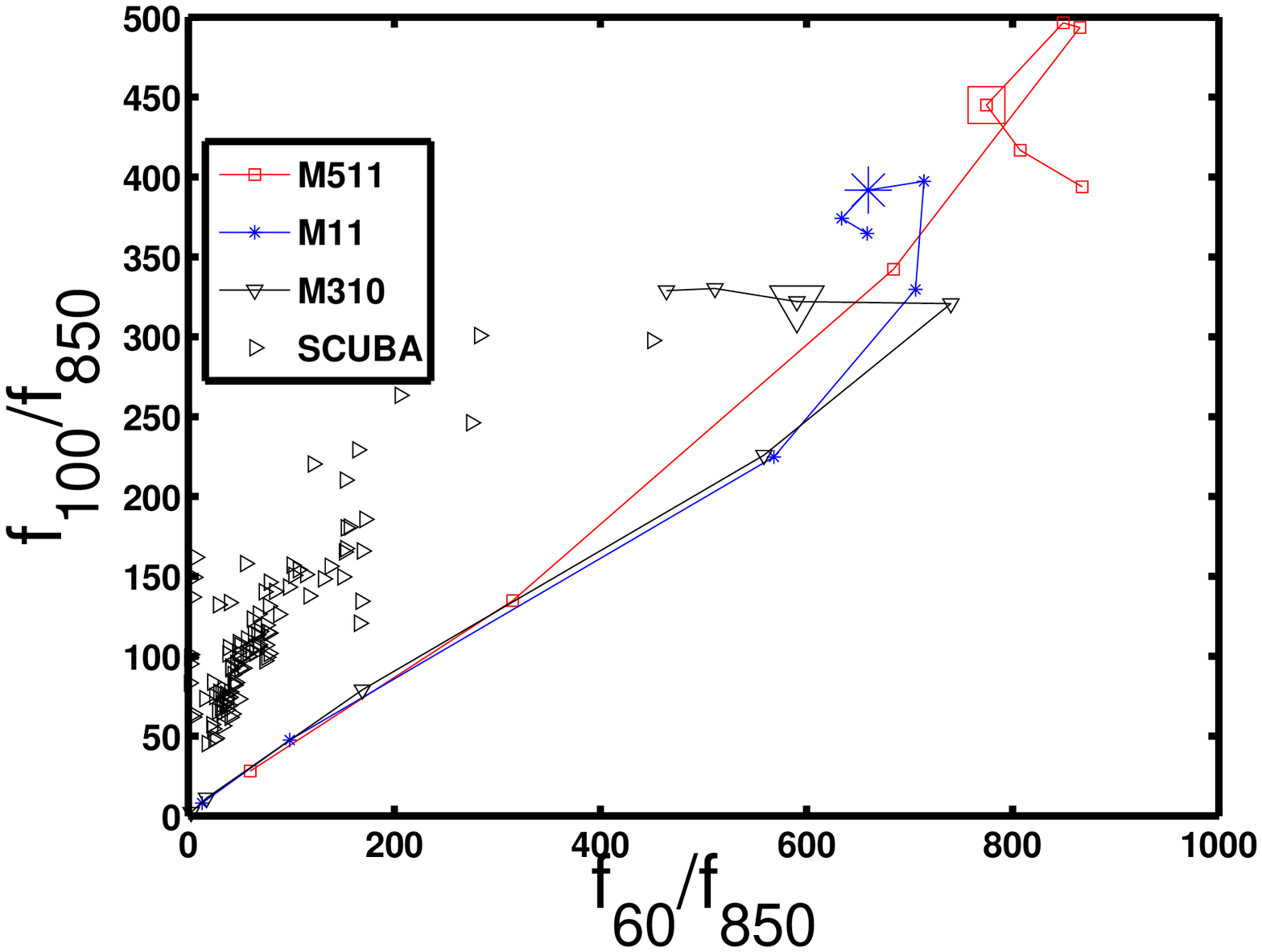}\\
\caption{Rest-frame $f_{100}/f_{850}$ flux density ratio vs  $f_{60}/f_{850}$ flux density ratio.  The data of SCUBA local galaxies
are from \cite{Dunne2000MNRAS.315..115D}. Lines are our model predictions of  $f_{100}/f_{850}$  vs $f_{60}/f_{850}$ evolution from 0.1 to 0.7 Gyr.
The large triangle, asterisk and square  are corresponding  for M310, M11 and M510 model predictions at 0.5Gyr, respectively. The high redshift galaxies and local SMGs have different FIR properties.}\label{60_100_850}
\end{center}
\end{figure}

The dust temperatures  are merely those corresponding to the best-fitting single temperature grey-body model, and are not the actual dust temperature.
We use the rest-frame 100-to-850 $\mu$m flux density ratio  versus 60-to-850 $\mu$m flux density ratio,  $f_{100}/f_{850}$ vs $f_{60}/f_{850}$, as a proxy to FIR properties of galaxies. Our predictions  of  M310, M11, M511 models from 0.1 Gyr to 0.7 Gyr are compared with  the  data of  SCUBA local  (recession velocity $\sim$ few thousand $ km \, s^{-1}$ )  galaxies \citep{Dunne2000MNRAS.315..115D} in Fig. \ref{60_100_850}.
Compared to SCUBA local   galaxies,   the young high redshift starburst galaxies in our models have  much higher $f_{60}/f_{850}$ given the same $f_{100}/f_{850}$.  The more massive galaxies have higher  $f_{100}/f_{850}$ and $f_{60}/f_{850}$.
It is clear that the SCUBA local SMGs and young high redshift star-burst galaxies have different FIR properties.

We argue that  caution is needed when adopting the local empirical UV-IR relationships and templates for  high redshift $z \sim 3$ LBGs to understand the physical processes in detail.

\section{Conclusions}\label{summary}
In this paper, we modeled  the rest frame UV-radio SED of the  stacked  MIPS-LBG at $z \sim 3$ by a new ``fitting" approach. In this self-consistent approach,  we derived the average galactic-wide properties of  MIPS-LBGs at $z \sim 3$ considering  the  stacked  MIPS-LBG  as a proxy of all MIPS-LBGs at $z \sim 3$.   Our findings can be summarized as follows:

\begin{enumerate}
\item  The new ``fitting" approach, which combines the chemical evolution model and GRASIL, can reproduce the  rest frame UV-radio SED of the stacked  MIPS-LBG at $z \sim 3$.  This approach suggests that the  MIPS-LBGs at $z \sim 3$ are likely young (0.3-0.6 Gyr) massive ($\sim 10^{11} M_{\odot}$) elliptical galaxies with a fast star-burst regime of star formation. Chemical enrichment and dust evolution history are provided by  chemical evolution models, calibrated at both low and high redshifts.

\item  We estimated that the  the average stellar mass of MIPS-LBGs is in the range of  $\sim 6\times 10^{10} M_{\odot}$  to $\sim 1\times 10^{11} M_{\odot}$. The ``best-fitting" stellar mass and SFR of  the stacked  MIPS-LBG are $8\times 10^{10} M_{\odot}$ and $ 200 M_{\odot}/yr$, respectively. The stellar mass is in agreement with  estimates done in other works and based on UV-optical SED fitting. This confirms that the derived stellar mass is  robust in SED fitting.

\item  The ``best-fitting" model of the  stacked  MIPS-LBG is the 0.5 Gyr $10^{11} M_{\odot}$ elliptical galaxy, which has dust mass of $M_{d}=7\times10^{7}M_{\odot}$ and gas mass of $M_{g}=1.6\times 10^{10}M_{\odot}$. Our ``best-fitting" dust mass of the stacked  MIPS-LBG is about a factor of eight less than the value based on  single temperature grey-body fitting \citep[$M_d = 5.5 \pm 1.6 \times 10^{8}M_{\odot}$ in][]{Rigopoulou2010MNRAS.tmpL.154R}.

\item The parameters of molecular clouds and dust of the Milky Way cannot fit the stacked MIPS-LBG SED.  Our ``best-fitting"  parameters reflect that:  i) MCs of  the stacked MIPS-LBG are more dense dusty environments than  in the Milky Way; ii) the dust size distributions in  stacked MIPS-LBG may be flatter than in the Milky Way; iii)  both small  and big size dust are needed to reproduce the  stacked MIPS-LBG SED; iv) non-negligible PAH also make an important contribution to MIR flux. More observational data, such as chemical abundances and emission lines,  are needed to make a more realistic  prediction  of the properties of all MIPS-LBGs, especially for dust intrinsic properties.

\item  We suggest that high redshifts star-forming ellipticals make a significant contribution to the population of ULIRGs.  The dust properties and morphology of high redshift star-burst galaxies could be different from local star-burst galaxies. We argue that  self-consistent SED models are needed to investigate the detailed properties of high redshift LBGs .

\end{enumerate}

\section*{Acknowledgments}
We thank the anonymous referee for valuable comments and useful suggestions which improved this work very much.  We  thanks G.E. Magids for providing the stacked MIPS-LBGs data and T. Bschorr for useful comments. X.L. Fan also thanks L. Silva and G. Granato for proving GRASIL and for many useful discussions.   F. Matteucci acknowleges financial support from PRIN MIUR2010-2011, Project "The Chemical and Dynamical Evolution of the Milky Way and Local Group Galaxies", prot. N.2010LY5N2T.  X.L.  Fan  acknowleges financial support from NSFC  (Grant No 11243006).

\bibliographystyle{apj}

\end{document}